\documentclass[10pt,twocolumn,groupedaddress,showpacs]{revtex4-1}

\usepackage{graphicx}
\usepackage{amsmath}

\newcommand{\ket}[1]{|#1\rangle}

\begin{document}

\title{Dominance of quantum over classical correlations: entropic and geometric approach}

\author{Z.~Walczak}
\email[]{z.walczak@merlin.phys.uni.lodz.pl}
\author{I.~Wintrowicz}
\author{K.~Zakrzewska}
\affiliation{Department of Theoretical Physics, University of Lodz,
Pomorska 149/153, 90-236 {\L}\'od\'z, Poland}

\date{\today}

\begin{abstract}
Recently, it has been shown that there exist quantum states for which quantum correlations dominate over classical correlations. 
Inspired by this observation, we investigate the problem of quantum correlations dominance for two-qubit Bell diagonal states in the Ollivier--Zurek paradigm, using both entropic and geometric approach to quantification of classical and quantum correlations.
In particular, we estimate numerically the amount of two-qubit Bell diagonal states for which quantum correlations dominate over classical correlations, and vice versa. 
Moreover, we show that in general these two approaches to quantification of correlations provide ambiguous results.
\end{abstract}


\maketitle

\section{Introduction}
In quantum information science, the problem of quantification of correlations present in quantum systems has been intensively studied both theoretically and experimentally for over two decades 
(for review, see \cite{Horodeccy09, Guhne09, Modi12}).
The most significant progress in this field has been achieved in the framework of paradigm based on 
the entanglement-separability dichotomy introduced by Werner \cite{Werner89}.

In the Werner paradigm, the quantum entanglement which is quantified by a vast number of different entanglement measures is the only type of quantum correlations that cannot exist without classical correlations (an opposite phenomenon has been reported recently for multiqubit states \cite{Kaszlikowski08}, however this result was questioned \cite{Walczak10, Bennett11}) and moreover quantum correlations do not dominate over classical correlations.

However, it has become clear gradually that the Werner paradigm is too narrow and needs reconsideration because separable quantum states can have non-classical correlations beyond quantum entanglement 
(see, e.g. \cite{Knill98, Braunstein99, Bennett99, Biham04, Pankowski08}).

The first step in this direction was taken by Ollivier and Zurek \cite{Ollivier01} who introduced an entropic measure of non-classical correlations called quantum discord. 
After the discovery \cite{Datta08, Datta09} that non-classical correlations other than entanglement can be 
responsible for the quantum computational efficiency of deterministic quantum computation with one qubit \cite{Knill98}, quantum discord became a subject of extensive studies in different contexts \cite{Modi12}.

Since evaluation of quantum discord involves a complicated optimization procedure, a geometric measure of 
non-classical correlations called geometric quantum discord was proposed to avoid this problem \cite{Dakic10}.   
As in the case of quantum discord, geometric quantum discord has  been extensively studied in different contexts \cite{Modi12}.

Therefore, in the Ollivier--Zurek paradigm there exist two internally coherent approaches to quantification of classical and quantum correlations present in bipartite quantum states.

Remarkably, in the Ollivier--Zurek paradigm quantum correlations may dominate over classical correlations,
contrary to the Werner paradigm, as has been shown recently in entropic approach to quantum and classical correlations.
In particular, this phenomenon was observed for a two-parameter family of two-qubit Bell diagonal states \cite{Luo08}, as well as for the $XXZ$ and $XX$ spin chains \cite{Sarandy09, Maziero10B}.

In this article, we investigate the problem of quantum correlations dominance for two-qubit Bell diagonal states in the Ollivier--Zurek paradigm, using both entropic and geometric approach.
In particular, we estimate numerically the amount of two-qubit Bell diagonal states with quantum correlations dominance and show that there exist two-qubit Bell diagonal states for which the two internally coherent approaches provide ambiguous results.

\section{Entropic approach to quantum and classical correlations}
 In quantum information theory, the quantum mutual information 
\begin{equation} 
\label{QMI}
I(\rho_{AB}) = S(\rho_{A}) + S(\rho_{B}) - S(\rho_{AB})
\end{equation}
quantifies the total correlations
present in a bipartite state $\rho_{AB}$, 
where $\rho_{A(B)}$ is the reduced state of the system $A(B)$ 
and $S(\rho) = - \text{Tr}(\rho \log_{2} \rho)$ is the von Neumann
entropy.
 
The quantum conditional entropy 
\begin{equation}
\label{QCE}
S(\rho_{B|A}) = S(\rho_{AB}) - S(\rho_{A})
\end{equation}
allows one to rewrite the quantum mutual information 
in the following form 
\begin{equation}
\label{QMI2}
I(\rho_{AB}) = S(\rho_{B}) - S(\rho_{B|A}).
\end{equation}  
The fact that the quantum conditional entropy quantifies 
the ignorance about the system $B$ that remains if we perform
measurement on the system $A$ allows one to find alternative 
expressions for the quantum conditional entropy and  
the quantum mutual information.

If the von Neumann projective measurement, 
described by a complete set of one-dimensional
orthogonal projectors, $\{\Pi_{i}^{A}\}$, corresponding 
to outcomes $i$, is performed, then the post-measurement state of the system $B$ 
is given by 
\begin{equation}
\label{B|i}
\rho_{B|i} = 
\text{Tr}_{A}[(\Pi_{i}^{A} \otimes I) \rho_{AB}
              (\Pi_{i}^{A} \otimes I)]/p_{i}^{A}
\end{equation}  
where $p_{i}^{A} = \text{Tr}[(\Pi_{i}^{A} \otimes I) \rho_{AB}]$.
The von Neumann entropies $S(\rho_{B|i})$, weighted by probabilities 
$p_{i}^{A}$, yield to the quantum conditional entropy of the system $B$ given the von Neumann projective measurement on the system $A$  
\begin{equation}
S_{\{\Pi_{i}^{A}\}}(\rho_{B|A}) = \sum_{i} p_{i}^{A} S(\rho_{B|i}),
\end{equation} 
and thereby the quantum mutual information, induced by this measurement, is defined by
\begin{equation}
J_{\{\Pi_{i}^{A}\}}(\rho_{AB}) = 
S(\rho_{B}) - S_{\{\Pi_{i}^{A}\}}(\rho_{B|A}).
\end{equation} 
The measurement independent quantum mutual information, 
interpreted as a measure of classical correlations present in a bipartite state $\rho_{AB}$ \cite{Henderson01, Ollivier01}, is defined  by  
\begin{align}
\label{CA}
C(\rho_{AB}) = \max_{\{\Pi_{i}^{A}\}} 
J_{\{\Pi_{i}^{A}\}}(\rho_{AB}). 
\end{align}
In general case,  $I(\rho_{AB})$ and $C(\rho_{AB})$ may differ and the difference, interpreted in a natural way as a measure of quantum correlations, is called quantum discord \cite{Ollivier01} 
\begin{align}
\label{DA}
D(\rho_{AB}) = I(\rho_{AB}) - C(\rho_{AB}).
\end{align}
The quantum discord $D(\rho_{AB})$ can be seen as the minimal amount of correlations which are lost when the non-selective von Neumann projective measurement is  performed on the system $A$ \cite{Luo10, Okrasa11}. Moreover, the quantum discord $D(\rho_{AB})$ is a lower bound for the global quantum correlations present in a bipartite state 
$\rho_{AB}$ \cite{Okrasa11}.

Since evaluation of quantum discord involves a complicated optimization procedure, the analytical expressions for quantum discord are known only for two-qubit Bell diagonal states \cite{Luo08}, 
for seven-parameter two-qubit $X$ states \cite{Ali10} 
(not always correct exactly, but approximately correct with a very small absolute error \cite{Huang13}), 
for two-mode Gaussian states \cite{Giorda10, Adesso10},
for a class of two-qubit states with parallel nonzero Bloch
vectors \cite{Li11B} and for two-qudit Werner and isotropic states 
\cite{Chitambar12}. 
Despite this fact, quantum discord has been studied in different contexts \cite{Modi12}, for example such as complete positivity of reduced quantum dynamics \cite{Rodriguez08, Shabani09}, 
broadcasting of quantum states \cite{Piani08},
random quantum states \cite{Ferraro10},
dynamics of quantum discord under both Markovian and non-Markovian evolution
\cite{Werlang09, Maziero09, Fanchini10,  Maziero10, Wang10, Hu11, Franco12}, 
operational interpretation of quantum discord \cite{Madhok11, Cavalcanti11}, 
connection between quantum discord and entanglement irreversibility \cite{Fanchini11B},
relation between quantum discord and distillable entanglement \cite{Streltsov11},
relation between quantum discord and distributed entanglement \cite{Fanchini11, Fanchini11C, Tan12},
interplay between quantum discord and quantum entanglement \cite{Luo08, Ali10, Qasimi11, Fanchini11D, Campbell12}, 
and monogamy of quantum discord \cite{Giorgi11, Prabhu12}.
Moreover, recently it has been shown that computing quantum discord is an NP-complete problem \cite{Huang14}.

\section{Geometric approach to quantum and classical correlations}
Because in the entropic approach it is difficult to compute correlations present in a bipartite state $\rho_{AB}$, 
even for a general two-qubit state, an alternative approach to this issue was proposed,   
where different types of correlations are quantified by a distance from a given state $\rho_{AB}$ to the closest state which does not have the desired property \cite{Modi10}.

In geometric approach, quantum correlations are quantified by a distance from the state $\rho_{AB}$ to the closest zero-discord state $\chi_{AB}$ and classical correlations are quantified by a distance from the state $\chi_{AB}$ to the closest product state $\pi_{AB}$. 
Of course, the amount of correlations present in a bipartite state $\rho_{AB}$ is determined by the choice of distance measure for quantum states.

Geometric quantum discord \cite{Dakic10}, the first measure of quantum correlations in the geometric approach, 
is based on the Hilbert--Schmidt distance.
Because geometric quantum discord involves a simpler optimization procedure than quantum discord, the analytical expression for geometric quantum discord was obtained for arbitrary  two-qubit states \cite{Dakic10},
as well as for arbitrary qubit-qudit states \cite{Hassan12, Rana12B}. 
As in the case of quantum discord, geometric quantum discord was studied in different contexts \cite{Modi12}, for example such as the quantum computational efficiency of deterministic quantum computation with one qubit \cite{Dakic10, Passante12}, dynamics of the geometric quantum discord  
\cite{Lu10, Altintas10, Yeo10, Li11, Bellomo12A, Bellomo12B}, relation between the geometric quantum discord and other measures of non-classical correlations \cite{Batle11, Qasimi11, Girolami11, Girolami11B, Rana12A, Okrasa12}.

However, recently it has been shown that contrary to quantum discord, geometric quantum discord is not a bona fide measure of quantum correlations, as has been shown explicitly in \cite{Piani12}, 
because of the lack of contractivity of the Hilbert--Schmidt norm, 
adopted as a distance measure, under trace-preserving quantum channels \cite{Hu13}.

Taking into account that the Hilbert--Schmidt norm is a special case of the Schatten $p$-norm (for $p=2$), 
geometric quantum discord based on the Schatten $p$-norm, adopted as a distance measure for quantum states,
was introduced 
\begin{align}
\label{GDA}D_{p}(\rho_{AB}) = \min_{\chi_{AB}} || \rho_{AB} - \chi_{AB} ||_{p}^{p},
\end{align}
where the minimum is over all zero-discord states $\chi_{AB}$,  
$|| X ||_{p} = [\text{Tr}((X^{\dagger} X)^{p/2})]^{1/p}$ is the Schatten $p$-norm 
and $p$ is a positive integer number \cite{Debarba12}.

Recently, it has been shown that only the Schatten $1$-norm is contractive under trace-preserving quantum channels \cite{Paula13}. Therefore in the geometric approach, quantum correlations present in a bipartite state $\rho_{AB}$ are quantified by   
\begin{align}
\label{QTR}
D_{1}(\rho_{AB}) = \min_{\chi_{AB}} || \rho_{AB} - \chi_{AB} ||_{1},
\end{align}
where $|| X ||_{1} = \text{Tr}[\sqrt{X^{\dagger} X}]$ is the trace norm \cite{Paula13},
and classical correlations present in a bipartite state $\rho_{AB}$ are quantified by 
\begin{align}
\label{CTR}
C_{1}(\rho_{AB}) = \min_{\pi_{AB}} || \chi_{AB} - \pi_{AB} ||_{1},
\end{align}
where the minimum is over all product states $\pi_{AB}$ \cite{Aaronson13}. 
Of course, $D_{1}(\rho_{AB})$ and $C_{1}(\rho_{AB})$ are defined up to a multiplicative constant 
(see e.g., \cite{Aaronson13, Ciccarello14}).

The analytical expression for geometric quantum discord $D_{1}(\rho_{AB})$ has been obtained for two-qubit Bell diagonal states \cite{Paula13, Nakano13} and for two-qubit $X$ states \cite{Ciccarello14}. 
Moreover, dynamics of geometric quantum discord $D_{1}(\rho_{AB})$ under Markovian evolution
has been studied both theoretically \citep{Aaronson13, Montealegre13} and experimentally \cite{Paula13B}.

\section{Dominance of quantum correlations over classical correlations for two-qubit Bell diagonal states} 
Two-qubit Bell diagonal states have the following form \cite{Horodecki96}
\begin{align}
\label{BDS}
\rho_{AB} = \frac{1}{4}(I \otimes I + 
\sum_{i=1}^{3} c_{i}\, \sigma_{i} \otimes \sigma_{i}),
\end{align}
where matrices $\sigma_{i}$ are the Pauli spin matrices 
and real numbers $c_{i}$ fulfill four conditions
\begin{subequations}
\begin{align}
& 0 \leq \frac{1}{4} (1 - c_{1} \mp c_{2} \mp c_{3}) \leq 1, \\
& 0 \leq \frac{1}{4} (1 + c_{1} \mp c_{2} \pm c_{3}) \leq 1.
\end{align}
\end{subequations} 
The above inequalities describe a tetrahedron with vertices
$(\pm 1,\pm 1,-1)$ and  $(\pm 1,\mp 1,1)$ 
corresponding to Bell states $\ket{\psi^{\pm}} = \frac{1}{\sqrt{2}}(\ket{01} \pm \ket{10})$ and
$\ket{\phi^{\pm}} = \frac{1}{\sqrt{2}}(\ket{00} \pm \ket{11})$, respectively \cite{Horodecki96}.

\begin{figure}[t]
   \centering
   \includegraphics[width=0.49\textwidth]{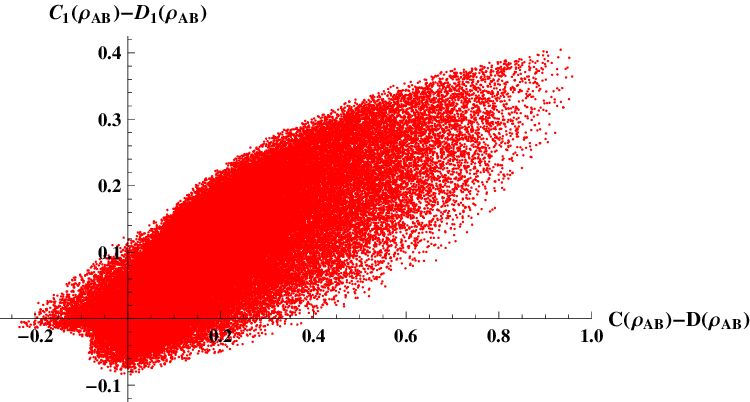}
    \caption{(Color online) Plot of $C_{1}(\rho_{AB}) - D_{1}(\rho_{AB})$ versus $C(\rho_{AB}) - D(\rho_{AB})$ 
    for $10^5$ random two-qubit Bell diagonal states.} 
   \label{fig1}
\end{figure}

In the entropic approach, classical and quantum correlations present in two-qubit Bell diagonal states are quantified  by \cite{Luo08}
\begin{subequations}
\label{EM}
\begin{align}
C(\rho_{AB}) & = 
\frac{1}{2} [ (1 - \alpha) \log_{2} (1 - \alpha) + (1 + \alpha) \log_{2} (1 + \alpha)],\\
D(\rho_{AB}) & = \frac{1}{4} 
[(1 - c_{1} - c_{2} - c_{3}) \log_{2}(1 - c_{1} - c_{2} - c_{3})
\nonumber\\
& + (1 - c_{1} + c_{2} + c_{3}) \log_{2} (1 - c_{1} + c_{2} + c_{3})
\nonumber\\
& + (1 + c_{1} - c_{2} + c_{3}) \log_{2} (1 + c_{1} - c_{2} + c_{3})
\nonumber\\
& + (1 + c_{1} + c_{2} - c_{3}) \log_{2} (1 + c_{1} + c_{2} - c_{3})]
\nonumber\\
& - \frac{1}{2} [ (1 - \alpha) \log_{2} (1 - \alpha) + (1 + \alpha) \log_{2} (1 + \alpha)], 
\end{align}
\end{subequations}   
where $\alpha$ is the maximum value of $|c_{1}|$, $|c_{2}|$ and $|c_{3}|$.

And in the geometric approach, the amount of classical and quantum correlations present in two-qubit Bell diagonal states are quantified  by \cite{Aaronson13}
\begin{subequations}
\label{GM}
\begin{align}
C_{1}(\rho_{AB}) & = -1 + \sqrt{1 + \alpha}, \\
D_{1}(\rho_{AB}) & = \frac{1}{2} \beta, 
\end{align}
\end{subequations}
where $\beta$ is the intermediate value of $|c_{1}|$, $|c_{2}|$ and $|c_{3}|$.

Evaluation of classical and quantum correlations for ten samples of $10^{7}$ random two-qubit Bell diagonal states shows that for $(80.612 \pm 0.021)\%$ of these states quantum correlations do not dominate over classical correlations,
i.e. for these states $C(\rho_{AB}) > D(\rho_{AB})$ and $C_{1}(\rho_{AB}) > D_{1}(\rho_{AB})$ 
(the first quarter of Fig.~\ref{fig1}).
But for $(5.294 \pm 0.008)\%$ of the random two-qubit Bell diagonal states quantum correlations dominate over classical correlations, 
i.e. for these states $C(\rho_{AB}) < D(\rho_{AB})$ and $C_{1}(\rho_{AB}) < D_{1}(\rho_{AB})$ 
(the third quarter of Fig.~\ref{fig1}).

Therefore, the two approaches to quantification of classical and quantum correlations provide unambiguous results for about $86\%$ of two-qubit Bell diagonal states.

However, for $(5.345 \pm 0.009)\%$ of the random two-qubit Bell diagonal states quantum correlations dominate over classical correlations in the entropic approach and vice versa in the geometric approach, 
i.e. for these states $C(\rho_{AB}) < D(\rho_{AB})$ and $C_{1}(\rho_{AB}) > D_{1}(\rho_{AB})$ 
(the second quarter of Fig.~\ref{fig1}),  
and for $(8.748 \pm 0.009)\%$ of the random two-qubit Bell diagonal states quantum correlations dominate over classical correlations in the geometric approach and vice versa in the entropic approach, 
i.e. for these states $C(\rho_{AB}) > D(\rho_{AB})$ and $C_{1}(\rho_{AB}) < D_{1}(\rho_{AB})$
(the fourth quarter of Fig.~\ref{fig1}).

Remarkably, the two internally coherent approaches to quantification of classical and quantum correlations 
provide ambiguous results concerning quantum correlations dominance for about $14\%$ of two-qubit 
Bell diagonal states.

\begin{figure}[t]
   \centering
   \includegraphics[width=0.49\textwidth]{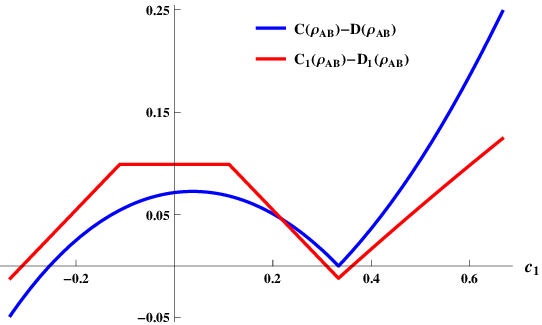}
    \caption{(Color online) Plot of $C(\rho_{AB}) - D(\rho_{AB})$ (blue line)
    and $C_{1}(\rho_{AB}) - D_{1}(\rho_{AB})$ (red line) 
    versus $c_{1}$ for two-qubit Bell diagonal states with 
    $-1/3 \leq c_{1} \leq 2/3$, $c_{2} = 1/9$ and  $c_{3} = -1/3$.} 
   \label{fig2}
\end{figure}

\section{Examples}
For two-qubit Bell diagonal states with 
\begin{align}
-1/3 \leq c_{1} \leq 2/3, \quad c_{2} = 1/9, \quad c_{3} = -1/3
\end{align}
the function $C(\rho_{AB}) - D(\rho_{AB})$ has two zeros (Fig.~\ref{fig2})
\begin{align}
a_{1} \simeq -0.2538596469, \quad a_{2} = 1/3,
\end{align}
and the function $C_{1}(\rho_{AB}) - D_{1}(\rho_{AB})$ has three zeros (Fig.~\ref{fig2})
\begin{align}
b_{1} = -4/\sqrt{3} + 2, \quad b_{1} = 4/\sqrt{3} - 2, \quad b_{3} = 13/36,
\end{align}
where $b_{1} < a_{1} < b_{2} < a_{2} < b_{3}$.

\subsection{Example 1}
For two-qubit Bell diagonal states with 
\begin{align}
b_{3} < c_{1} \leq 2/3, \quad c_{2} = 1/9, \quad c_{3} = -1/3
\end{align}
quantum correlations do not dominate over classical correlations,
i.e. for these states $C(\rho_{AB}) > D(\rho_{AB})$ and $C_{1}(\rho_{AB}) > D_{1}(\rho_{AB})$.
\subsection{Example 2}
For two-qubit Bell diagonal states with 
\begin{align}
-1/3 \leq c_{1} < b_{1}, \quad c_{2} = 1/9, \quad c_{3} = -1/3
\end{align}
quantum correlations dominate over classical correlations, 
i.e. for these states $C(\rho_{AB}) < D(\rho_{AB})$ and $C_{1}(\rho_{AB}) < D_{1}(\rho_{AB})$.
\subsection{Example 3}
For two-qubit Bell diagonal states with 
\begin{align}
b_{1} < c_{1} < a_{1}, \quad c_{2} = 1/9, \quad c_{3} = -1/3
\end{align}
quantum correlations dominate over classical correlations in the entropic approach and vice versa in the geometric approach, 
i.e. for these states $C(\rho_{AB}) < D(\rho_{AB})$ and $C_{1}(\rho_{AB}) > D_{1}(\rho_{AB})$.
\subsection{Example 4}
For two-qubit Bell diagonal states with 
\begin{align}
a_{2} < c_{1} < b_{3}, \quad c_{2} = 1/9, \quad c_{3} = -1/3
\end{align}
quantum correlations dominate over classical correlations 
in the geometric approach and vice versa in the entropic approach, 
i.e. for these states $C(\rho_{AB}) > D(\rho_{AB})$ and $C_{1}(\rho_{AB}) < D_{1}(\rho_{AB})$.

\section{Summary}
We have investigated the problem of quantum correlations dominance for two-qubit Bell diagonal states in the entropic as well as the geometric approach to quantification of classical and quantum correlations.

In particular, we have shown numerically that these two approaches provide unambiguous results 
for about $86\%$ of two-qubit Bell diagonal states.
Namely, for about $81\%$ of these states quantum correlations do not dominate over classical correlations, 
but for about $5\%$ of these states quantum correlations dominate over classical correlations.

Moreover, we have discovered numerically that these two internally coherent approaches provide ambiguous results for about $14\%$ of two-qubit Bell diagonal states.
Namely, for about $5\%$ of the random two-qubit Bell diagonal states quantum correlations dominate over classical correlations in the entropic approach and vice versa in the geometric approach, 
but for about $9\%$ of these states quantum correlations dominate over classical correlations 
in the geometric approach and vice versa in the entropic approach. 
This remarkable phenomenon is important both from a theoretical and an experimental point of view, 
because it shows that two widely used approaches to quantification of classical and quantum correlations present in quantum states are not  fully equivalent and in general they provide ambiguous results.

\begin{acknowledgments}
This work was supported by the University of Lodz Grant, 
the Polish Ministry of Science and Higher Education 
Grant No. N N202 103738, and the Polish Research Network 
{\em Laboratory of Physical Foundations of Information Processing}. 
\end{acknowledgments}

\end{document}